\documentstyle{europhys}

\def\etal{{\hbox{{\tenit\ et al.\/}\tenrm :\ }}}

\def\And{{\rm and\ }}

\def\stars{\bigskip\centerline{***}\medskip}

\newif\ifboo \boofalse

\def\Review#1{\boofalse{\it #1},}
\def\Name#1{{\sc #1},}
\def\Vol#1{\ifboo Vol. {\bf #1}\else{\bf #1}\fi}
\def\Year#1{\ifboo #1\else(#1)\fi}
\def\Book#1{\bootrue{\it #1},}
\def\Page#1{\ifboo {\rm p. #1}\else{\rm #1}\fi}

\begin{document}
%
%
%
\euro{}{}{}{1998}
\Date{2 July 1998}
\shorttitle{S. J. Lee \etal GLASSY RELAXATION AND ... .}

\title {Glassy Relaxation and Breakdown of the Stokes-Einstein Relation
in the Two Dimensional Lattice Coulomb Gas of Fractional Charges}

\author{Sung Jong Lee\inst{1}, Bongsoo Kim\inst{2} \And Jong-Rim Lee\inst{3}}

\institute{
     \inst{1} Department of Physics - The University of Suwon, Hwasung-Gun, \\
      Kyunggi-Do 445-743, Korea \\
     \inst{2} Department of Physics - Changwon National University, \\
       Changwon 641-773, Korea \\
     \inst{2} NSSPC - College of Engineering, Changwon National University,\\
         Changwon 641-773, Korea }

%
%
%
%
%
%
\rec{}{}
%
%
%

\pacs{
\Pacs{64}{70Pf}{Glass transitions}
\Pacs{64}{60Cn}{Statistical Mechanics of model systems}
      }
\maketitle
%
%
%
\begin{abstract}
We present Monte Carlo simulation results on the 
equilibrium relaxation of the two dimensional lattice Coulomb gas with fractional
charges, which exhibits a close analogy to the primary relaxation of fragile 
supercooled liquids.  Single particle and collective relaxation dynamics show
that the Stokes-Einstein relation is violated  at low temperatures, 
which can be characterized by a fractional power law relation between the self-diffusion 
coefficient and the characteristic relaxation time. 
The microscopic spatially heterogeneous structure responsible for the violation is identified. 
\end{abstract}

%
%
%
%
%

The dynamics of supercooled liquids approaching the glass transition
remains one of the most fundamental problems in condensed matter physics \cite{review}.
An important question concerning transport properties of  
supercooled liquids is the validity and possible breakdown of 
the Stokes-Einstein (SE) relation. 
There have been many experimental results \cite{fujara} 
which show that highly supercooled liquids exhibit breakdown of the SE relation.
Although there exist some theoretical attempts \cite{stilinger},
the underlying microscopic mechanism for the violation of the SE relation 
is not well understood.

In this work, by probing single particle and 
collective relaxation dynamics of the two dimensional (2D) lattice Coulomb gas 
(CG)  with fractional charges parameterized by charge frustration $f=p/q$, we demonstrate that 
this {\em non-random} lattice model \cite{bm,kl} on a square lattice exhibits
equilibrium relaxation behavior that closely resembles the primary 
relaxation of fragile supercooled liquids. 
Moreover, we observe that the model exhibits a strong violation of 
the SE relation at low temperatures, which is characterized by 
a fractional power law relation between the self-diffusion coefficient 
and the characteristic relaxation time. 
Recent molecular dynamics simulations on the 2D  \cite{ph,nc} and  3D binary liquids \cite{tm} 
also report the breakdown of the SE relation.

General 2D CG on a square lattice is described by the following Hamiltonian that 
can be mapped from uniformly frustrated XY (UFXY) model by means of 
Villain transformation \cite{villain},

\begin{equation}
{\cal H}_{CG}={1 \over 2}\sum_{ij}q_iG(r_{ij})q_j \label{eq:Hcg}
\end{equation}
where $r_{ij}$ is the distance between the sites $i$ and $j$, and  
the magnitude of charge $q_i$ at site $i$ can take either $1-f$ or $-f$ at lowest excitation, 
where $f$ is called the frustration parameter. 
Charge neutrality condition $\sum_iq_i=0$ implies that the number density of 
the positive charges is equal to $f$. 
We thus can  view the system as a lattice gas of charges of unit magnitude 
upon uniform negative background charges.  
The lattice Green's function $G(r_{ij})$  solves the equation
$\Delta^2 G(r_{ij})=-2\pi\delta_{r_{ij},0}$, where $\Delta^2$ is 
the discrete lattice Laplacian. 
For large separation $r$, one gets $G(\vec r)\simeq -\ln r$ \cite{jrl}. 

Here, we are concerned with the case of fractional charges with $f$ approaching 
an irrational value $f=1-g$ where $g$ ¤Ôis the golden mean  $g=(\sqrt{5}-1)/2$.
While this model system has no intrinsic disorder,  
strong incommensurate nature of low temperature configurations induces 
local frustration of the charges.  
The nature of the equilibrium phase at low temperature of this CG was 
clarified in a recent extensive Monte Carlo (MC) simulations
carried out by Gupta, Teitel, and Gingras (GTG) \cite{gtg}.
They showed that in contrast to Halsey's earlier result \cite{halsey}, the system undergoes 
a first order phase transition to {\em ordered} structures at a finite temperature.
This ordered structure is characterized by diagonal stripes that are
alternatingly completely filled, completely empty, and partially filled with 
charges (see Fig.~1 of Ref.[6]). 

In our MC simulations, the initial disordered random configuration 
is updated according to the standard Metropolis algorithm
by exchanging the charges at randomly selected nearest neighbor (NN) or
next nearest neighbor (NNN) pairs as was done by GTG \cite{gtg}.  
We find that this NNN hopping algorithm is particularly
effective in terms of simulation time as compared with NN hopping alone, as was
emphasized by GTG.  The presented results are averages over $100 \sim 300$
different random initial configurations depending on the temperature.
As for the values of fractional charge parameter $f=p/q$, we selected various 
low order Fibonacci approximants to $f=1-g$, 
such as $f=3/8$, $5/13$, $8/21$, $13/34$ on a square 
lattice of linear sizes $N=32$, $N=39$,  $N=42$, and $N=34$, respectively,  with periodic
boundary conditions. The presented data below are those for $f=13/34$. We found that 
qualitative features of relaxation dynamics are the same for other values of the frustration $f$
given above. 

In order to probe the equilibrium relaxation behavior of the system, 
we measured the on-site charge density autocorrelation functions, 

\begin{equation}
C(t)=\langle \sum_{i=1}^{N^2} q_i(t_w)q_i(t_w+t) \rangle / N^2,
\end{equation}
where $t_w$ is the waiting time, which is taken well after the 
equilibration of the system at a given temperature (by probing the energy relaxation),
and the bracket $< \cdots >$ represents an average over different random 
initial configurations. 
The self-diffusion property is measured by the mean square displacement of 
a single particle 

\begin{equation}
<(\Delta \vec{r} )^2 > 
=\langle \sum_{i=1}^{N^2} (\vec{r}_i(t_w+t)-\vec{r}_i(t_w))^2 \rangle / N^2,
\end{equation}
where $\vec{r}_i(t)$ is the position vector of the charge at site $i$ at time $t$.

Shown in Fig.~1 is the on-site charge autocorrelation 
function $C(t)$ for temperatures from $T=0.1$ down to $T=0.028$.
From this figure, we observe a dramatic slowing down over about six 
decades of time scale in the structural relaxation for this temperature range. 
One can extract a characteristic time scale $\tau(T)$ which, for example,  is
defined as $C(t=\tau(T))=1/2$ for each temperature $T$. 
As the inset of Fig.~1 clearly shows, the temperature dependence of 
the relaxation time exhibits a non-Arrhenius behavior. 
A good Vogel-Tamman-Fulcher fit 
$\tau(T)=\tau_0 \exp(D_0 T_0/(T-T_0))$ is obtained for $T_0 \simeq 0.016$, 
$\tau_0 \simeq 1.66$, and $D_0 \simeq 6.40$.  
The value of $D_0$ indicates that the charge relaxation is similar to 
the primary relaxation in a typical fragile liquids \cite{torell}.
The time dependence of the relaxation function is characterized by 
a power law relaxation $C(t)=1-A t^{b(T)}$ (known as the von Schweidler relaxation) 
in the early time regime and a stretched exponential relaxation 
$C(t)=C_0 \exp(-A't^{\beta(T)})$ in the late time regime. 
The two exponents $b$ and $\beta$ are observed to be temperature dependent at
low temperatures, which gives rise to breakdown of the time-temperature superposition
of the relaxation.  The detailed analysis for the temperature dependence of the two exponents
and the scaling property of the relaxation function will be given elsewhere \footnote{
Work in preparation}.
These results clearly indicate that the equilibrium relaxation
in the 2D CG with fractional charges closely resembles the primary relaxation of
typical fragile supercooled liquids. 
Similar glassy relaxation in a densely frustrated XY model has recently been reported \cite{kl}.
According to GTG \cite{gtg}, the system undergoes a first order 
phase transition into an ordered phase at $T_m \simeq 0.03$. 
We note from Fig.~1 that the  relaxation dynamics  
does not show any distinctive feature above and below the melting temperature $T_m$.
It is interesting to note that the dynamic crossover 
$T_{cr} \simeq 0.045$ lies above the first order transition temperature $T_m \simeq 0.03$. 
Being defined on a lattice with incommensuration of the charge density with the underlying lattice, 
our model has inherent frustration at all temperatures with many metastable configurations.
This can be contrasted with the case of real supercooled liquids where frustrations are dynamically 
generated at low temperatures and the dynamic crossover occurs below the melting temperature.


Gross features of the single particle dynamics is described by the mean 
square displacement $ <(\Delta \vec{r} )^2 > $, which is shown in Fig.~2 for various temperatures.
$<(\Delta \vec{r} )^2 >$ exhibits an early time subdiffusive regime and crosses over into 
late time diffusive regime.   Early time subdiffusive behavior is thought to be 
coming from local frustrated motions of charges before reaching an average displacement
of unit lattice spacing.
Figure 3 shows an Arrhenius plot for the two time scales $D^{-1}$ and $\tau$ 
obtained from the single particle and collective relaxation dynamics, 
respectively. Here, while at high temperature side  the two time scales are 
proportional to each other, at low $T$ ($T < T_{cr} \simeq 0.045$), it is clearly
observed that proportionality between the two time scales ceases 
to hold due to the weaker temperature 
dependence of the diffusion coefficient, {\it i.e.}, increasing  enhancement of diffusion at 
lower temperatures.
This is quite analogous to the violation of the SE relation ($D=T/a \eta$, where
$a$ is a molecular length and $\eta$ is the viscosity of the liquid) observed 
in experiments on supercooled liquids  \cite{fujara} 
since the structural relaxation time $\tau$ can be considered as proportional to $\eta/T$.  
In the inset of Fig.~3 is shown a double-log plot of $D^{-1}$ versus $\tau$.
We find from this figure that above $T_{cr}$, $D^{-1} \sim \tau$, 
whereas below $T_{cr}$, a fractional SE relation holds with $D^{-1} 
\sim \tau^{\theta}$ with $\theta \simeq 0.67$.
This can be considered as a strong breakdown of the standard SE relation 
when we compare our value 0.67 with one experimental result \cite{fujara} on a fragile 
liquid orthoterphenyl giving $\theta_{OTP} \simeq 0.75$.  

%
 
In order to understand the underlying microscopic mechanism of the intriguing 
separation of the two time scales $D^{-1}$ and $\tau$, we looked into the typical charge 
configurations  
of the system at low temperature and also the characteristic features of 
charge motions.   Typical charge configurations, as shown in Fig.~4, exhibit local 
ordered domains 
and interfacial regions due to mismatch between adjacent domains. 
For a fixed quenching temperature, the average size of these local domains  
reaches a certain length scale when the system equilibrates.
After equilibration, the system structurally rearranges 
itself going from one configuration to another with local domains of similar length scale.
%
%
Figure 5 shows the trajectories of moving positive charges over a time interval of 
700 MC steps for $T=0.03$ (corresponding to Fig. 4(b)). 
We can see that there exists local regions with actively moving charges and other 
region with relatively inert regions.
Among the active regions, we can find those charges moving along partially filled 
diagonal channels over a long distance. We also find some extended interfacial 
regions where no discernible local order can be identified, that exhibit a sort
of fluidized motions.
Enhancement of particle diffusion is most probably due to the motions of charges
along the partially filled diagonals as well as those fluidized motions in 
the extended interfacial regions. 
These fastly moving regions in  surroundings of  very slowly moving regions offer a specific
example for spatial heterogeneity in  glassy systems \cite{onuki,harrowell}, 
which was often thought of as the physical mechanism for breakdown of the 
SE relation. It would be important to  provide further quantitative
measure of spatial heterogeneity in these systems.


There are a few important aspects that  need further consideration.
One of them relates to an important distinction between the present 2D CG and related
UFXY model. That is, the low temperature ordered structure for vortices
in the UFXY model, as Denniston and Tang \cite{dt} indicate, 
does not exhibit partially filled diagonals, but they consist of staircase structure or 
staircase structure with some regular defects depending on the values of $f$. 
Therefore detailed study on the UFXY model, in relation to the breakdown of the SE relation  
would be valuable. 

In summary, we have shown that the 2D lattice CG with dense charge frustration exhibits
a slow relaxation quite similar to the primary relaxation of typical fragile liquids, 
including a strong violation of the SE relation.
We have identified the microscopic heterogeneous structure which is responsible for
the violation of the SE relation at low temperatures.

We thank M. Y. Choi, P. Harrowell, K. Kawasaki, A. Onuki, S. Teitel, and M. Tokuyama for 
discussions.
This work was supported  in part by the BSRI (Grant No. BSRI 98-2412), 
Ministry of Education, Korea and by the Korea Research Foundation Grant No. 1997-003-D00086 (S.J.L.).

\stars
%

\bigskip

\bigskip

\noindent FIGURE CAPTIONS

\bigskip

\noindent Fig.1: The charge autocorrelation functions for temperatures 
$T = 0.1$, $0.08$, $0.06$, $0.05$, $0.04$, $0.035$, $0.032$, $0.03$, $0.028$. 
Inset: An Arrhenius plot for the characteristic relaxation time defined as
$C(\tau) =1/2$, where solid line is the best Vogel-Tamman-Fulcher fit.

\bigskip

\noindent Fig.2: Squared displacement  $<(\Delta \vec{r} )^2 >$ versus time $t$ 
for the same temperatures as in fig.~1.

\bigskip

\noindent Fig.3: Arrhenius plot for the two time scales $D^{-1}$ and $\tau$.
Inset: Double-log plot of $D^{-1}$ versus $\tau$. We can clearly see 
a crossover from high temperature regime with $D^{-1} \propto \tau$ 
to low temperature regime with breakdown of the Stokes-Einstein relation,
which can be characterized by $D^{-1} \propto \tau^{\theta}$, $\theta \simeq 0.67$.

\bigskip

\noindent Fig.4: Typical charge configurations at (a) $T=0.05$, and (b) $T=0.03$.
Positive charges are represented by filled squares.

\bigskip

\noindent Fig.5:
Trajectories of moving positive charges  in Fig.~4 (b) 
 over a time interval of $700$ MC steps. Arrows indicate 
the directions of single charge motions.




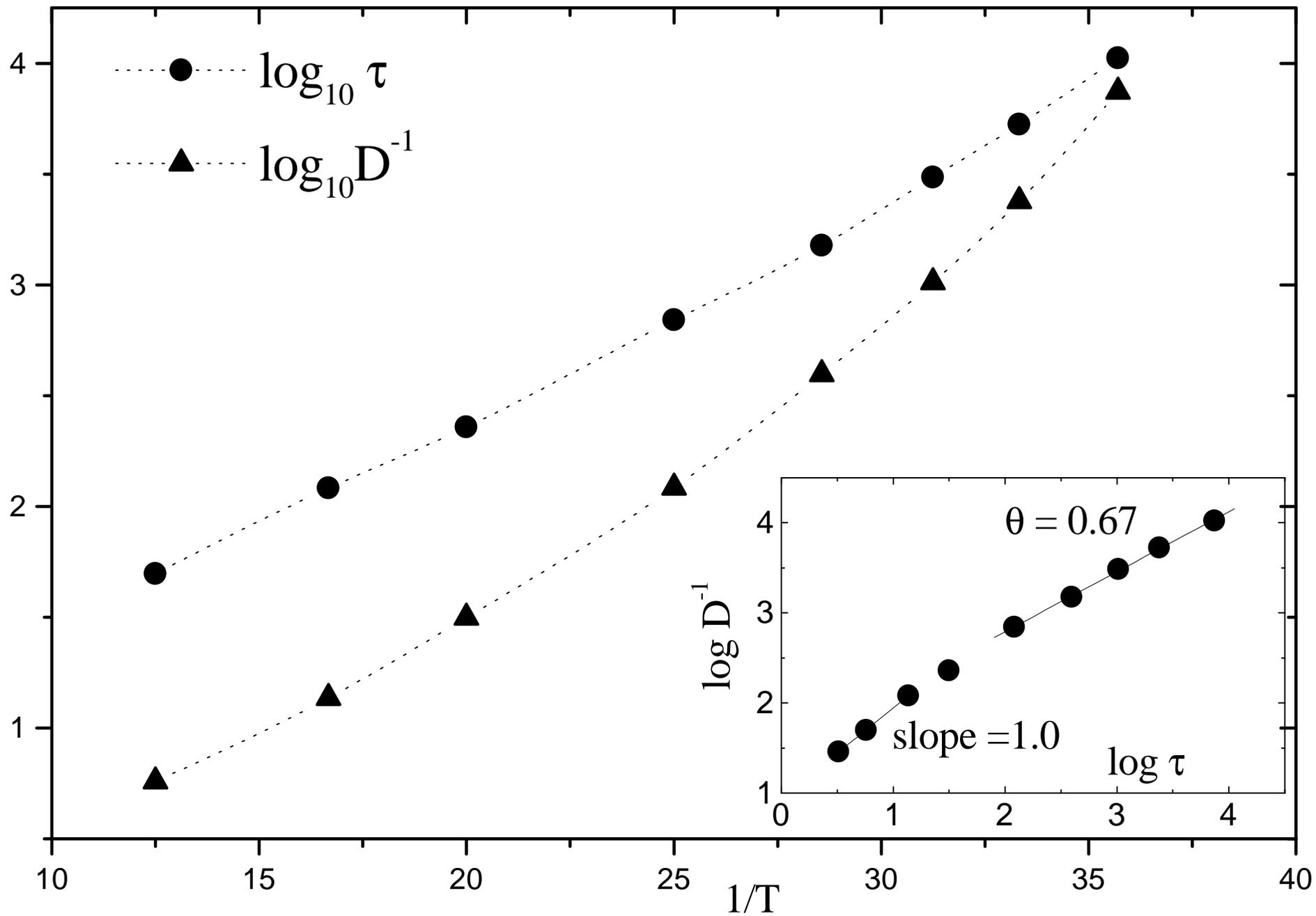
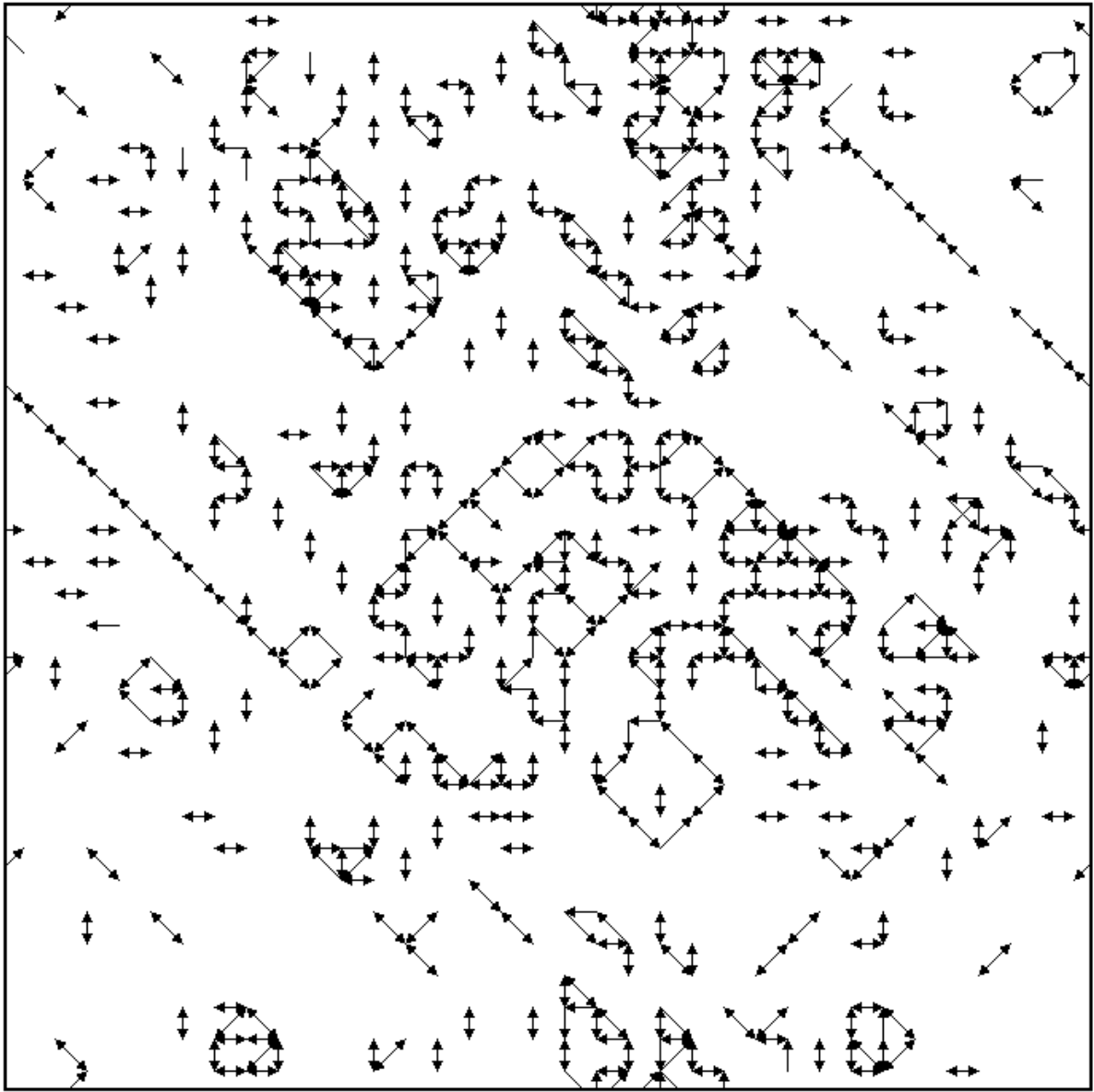
\end{document}